\documentclass[a4paper]{jpconf}
\usepackage{graphicx}
\begin{document}
\title{High-resolution molecular line observations of active galaxies}

\author{S~Garc\'{\i}a-Burillo$^1$, F~Combes$^2$, A~Usero$^{1,3}$ and J~Graci\'a-Carpio$^{1,4}$}

\address{$^1$ Observatorio de Madrid, OAN, Alfonso XII, 3, E-28014 Madrid, Spain}
\address{$^2$ Observatoire de Paris, LERMA, 61 Av. de l'Observatoire, F-75014 Paris,
France}
\address{$^3$ Centre for Astrophysics Research, University of Hertfordshire, College 
Lane, Hatfield AL10 9AB, UK}
\address{$^4$ Max-Planck Institut f\"r Extraterrestrische Physik (MPE), Giessenbachstrasse 1, D-85741 Garching, Germany}
\ead{s.gburillo@oan.es, francoise.combes@obspm.fr, a.usero@oan.es, j.gracia@oan.es}

\begin{abstract}
The study of the content, distribution and kinematics of interstellar gas is a key to understand the origin and maintenance of both starburst and nuclear (AGN) activity in galaxies. The processes involved in AGN fueling encompass a wide range of scales, both spatial and temporal, which have to be studied. Probing the gas flow from the outer disk down to the central engine of an AGN host, requires the use of specific tracers of the interstellar medium adapted to follow the change of phase of the gas as a function of radius. 
Current mm-interferometers can provide a sharp view of the distribution and kinematics of molecular gas in the circumnuclear disks of galaxies through extensive CO line mapping. As such, CO maps are an essential tool to study AGN feeding mechanisms in the local universe. This is the scientific driver of the {\it NUclei of GAlaxies} (NUGA) survey, whose latest results are here reviewed. On the other hand, the use of specific molecular tracers of the dense gas phase can probe the feedback influence of activity on the chemistry and energy
balance/redistribution in the interstellar medium of nearby galaxies. Millimeter interferometers are able to unveil the strong chemical differentiation present in the molecular gas disks of nearby starbursts and AGNs. Nearby active galaxies can be used as local templates to address the study of more distant galaxies where both star formation and AGN activity are deeply embedded. 

\end{abstract}

\section{The feeding of nuclear activity in galaxies}

The phenomenon of nuclear activity in galaxies is viewed as the result of the feeding of supermassive black holes. Active Galactic Nuclei (AGN) are fed with material lying originally in the disk of the galaxy, therefore far away from the gravitational influence of the black hole. This implies in practice that the gas has to lose nearly all of its angular momentum on its way to the central engine. A correlation has been found between the presence of $\sim$kpc-scale non-axisymmetric perturbations and the existence of activity in quasars, reckoned as high luminosity AGNs (HLAGNs). However, the search for a {\it universal} feeding mechanism in low luminosity AGNs (LLAGNs) had proved unsuccessful thus far \cite{kna00,com03,mart03,mart04}. The feeding problem is complicated by the fact that the AGN duty cycle ($\sim$10$^{6-7}$ years) might be shorter than the lifetime of the feeding mechanism itself \cite{wad04, kin07}. This time-scale conspiracy translates into a spatial-scale problem that can only be tackled by probing the critical scales for AGN feeding ($<$10--100~pc). These spatial scales can only be reached with current mm-interferometers in nearby LLAGNs, a similar study in HLAGNs awaits the advent of future large interferometers like the Atacama Large Millimeter Array (ALMA).  The gas masses involved in the feeding of HLAGNs, and possibly also the mechanisms of angular momentum transfer, could be different compared to those of LLAGNs \cite{sel04}. A study of AGN feeding in nearby LLAGNs can help us to address the fueling problem in general, however. In particular, these type of studies lay the ground for future surveys where the whole luminosity range of AGNs will be explored. As molecular gas represents the dominant phase of neutral gas in galactic nuclei, rotational lines of carbon monoxide (CO) are well suited to undertake high-spatial resolution ($\simeq$1$^{\prime\prime}$) interferometer mapping of the central kiloparsec disks of AGNs.

\subsection{The NUGA survey}

The NUclei of GAlaxies--NUGA--project, described by \cite{gb03a,gb03b}, is the first high-resolution ($\sim$0.5$^{\prime\prime}$--1$^{\prime\prime}$) $^{12}$CO survey of a sample of nearby ($D$=4--40~Mpc) LLAGNs including the full sequence of AGN activity types (Seyferts, LINERs and transition objects). Observations have been carried out with the IRAM Plateau de Bure Interferometer (PdBI).  The core sample of NUGA includes 12 targets for which we have the highest spatial resolution ($<$10--100~pc) and sensitivity images (3-$\sigma$--detection limit$\sim$a few 10$^{5-6}$M$_{\odot }$). This core sample is being expanded to reach a $\sim$25 galaxy sample by including data of other LLAGNs from the PdBI archive.  NUGA maps allow us to probe the gas flows at critical spatial scales where secondary modes embedded in the kpc-scale perturbations are expected to take over in the fueling process. To connect the large scales with the small scales, and have a complete picture of the fueling processes, a parallel HI survey has been conducted with the Very Large Array (VLA). This survey is used to probe the distribution and kinematics of neutral gas in the outer disks of NUGA galaxies \cite{haa08a, haa08b, gb08a}. 

One of the main results of NUGA is the identification of a wide range of morphologies in the central kpc-disks of these LLAGNs. This includes one-arm spirals or $m=1$ instabilities (NGC~4826: \cite{gb03b}; NGC~3718: \cite{kri05}), symmetric rings (NGC~7217: \cite{com04}; NGC~3147: \cite{cas08a}), gas bars and two-arm spirals (NGC~4569: \cite{boo07}; NGC~2782: \cite{hun08}; NGC~6574: \cite{lin08}; NGC~6951: \cite{gb05,kri07}), as well as mostly axisymmetric disks (NGC~5953: \cite{cas08b}). By itself, this result already suggests that several mechanisms, instead of one single universal mechanism, can cooperate to fuel the central engines of LLAGNs.


\begin{figure}[tb!]
\begin{center}
\includegraphics[width=\textwidth]{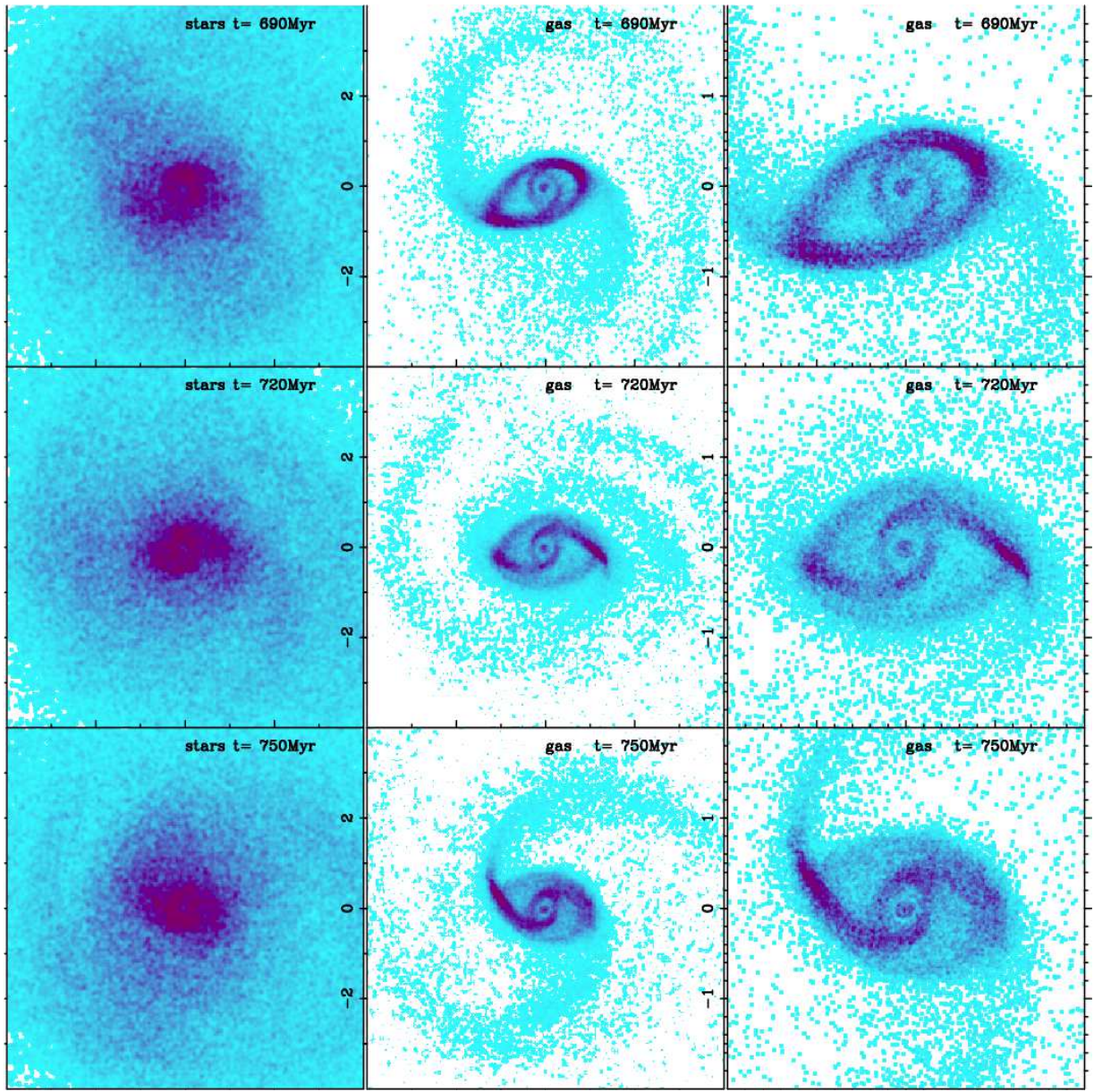}
\caption{\label{n2782} Snapshots of the stellar {\it (left panel)} and gas distributions {\it (mid and right panels)} at different epochs (in Myr) as seen in the numerical simulations that follow the evolution of a model galaxy to fit the distribution and kinematics of molecular gas in the NUGA galaxy NGC~2782 \cite{hun08}. Axes are in kpc. Through dynamical decoupling of the nuclear bar with respect to the outer bar, the gas falls progressively inward. The gas infalling along the ``dust lanes'' is aligned along the nuclear bar, where the CO is observed in NGC~2782. Figure adapted from \cite{hun08}.}
\end{center}
\end{figure}


\begin{figure}[pt!]
\begin{minipage}{17pc}
\includegraphics[width=17pc]{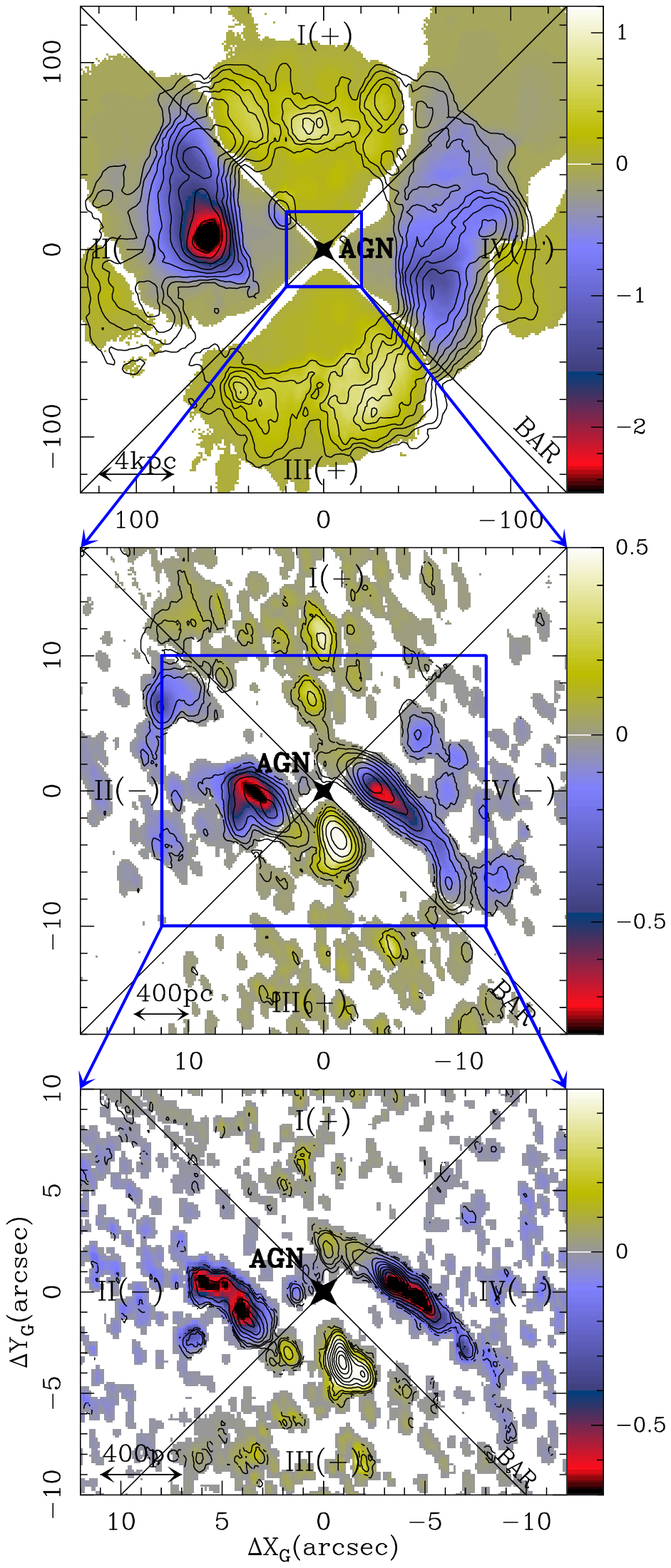}
\caption{\label{n4579-1}{\bf a)}({\it Upper panel})~We overlay the HI contours with the the gravity torque map of NGC~4579.  The derived torques change sign as expected if the {\it butterfly} diagram, defined by the orientation of quadrants I-to-IV, are due to the action of the large-scale bar. {\bf b)}({\it Mid panel})~ The same as {\bf a)} but for CO(1--0). {\bf c)}({\it Lower panel})~The same as {\bf a)} but for CO(2--1). Figure adapted from ~\cite{gb08a}.}
\end{minipage}\hspace{1.5pc}
\begin{minipage}{21.5pc}
\includegraphics[width=21.5pc]{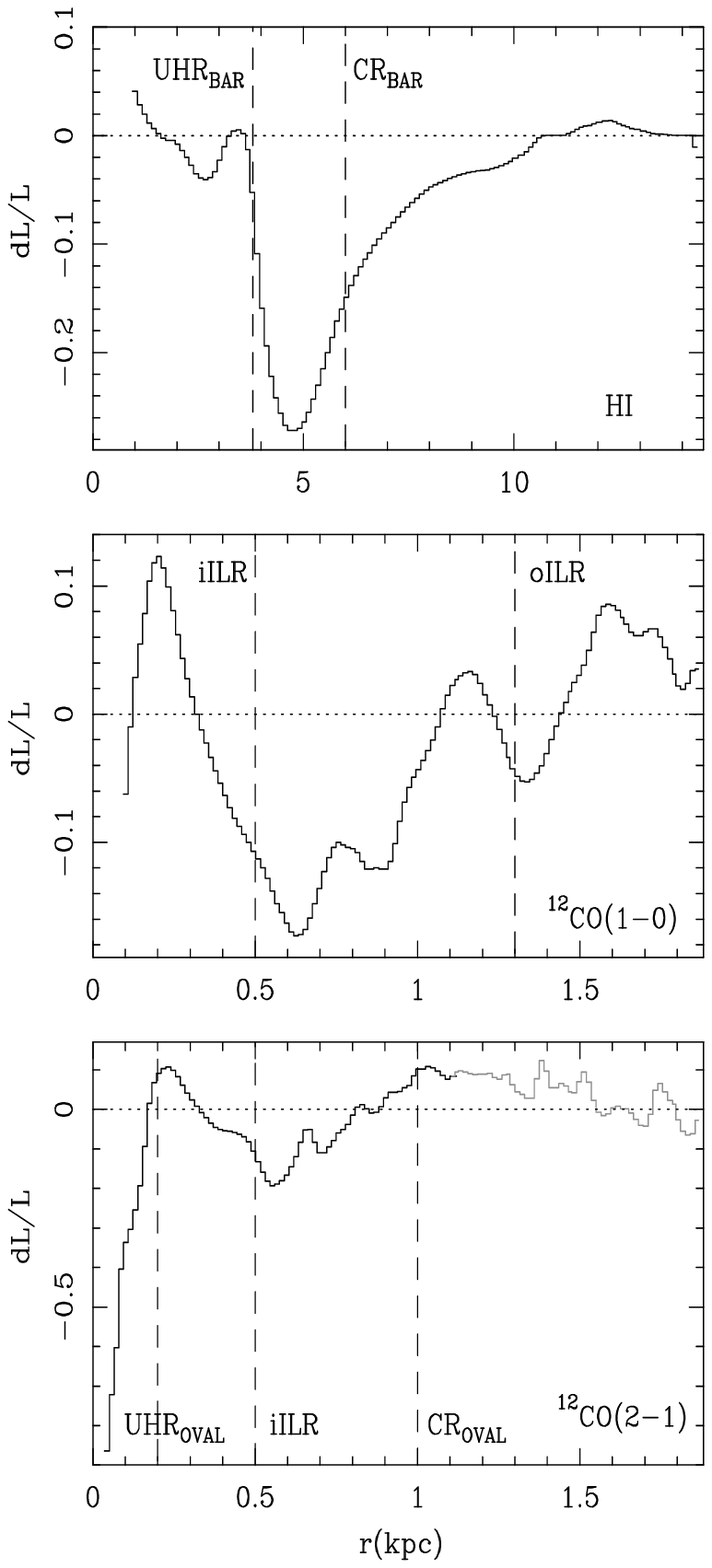}
\caption{\label{n4579-2}~The average fraction of the angular momentum transferred from/to the gas in one rotation ($dL/L$) are plotted as a function of radius, as derived from the HI ({\it upper panel}), CO(1-0) ({\it mid panel}) and CO(2-1) ({\it lower panel}) maps of the disk of NGC\,4579. The locations of BAR resonances (iILR, oILR, CR$_{BAR}$, UHR$_{BAR}$) and those of OVAL resonances (CR$_{OVAL}$, UHR$_{OVAL}$) are highlighted. Figure adapted from ~\cite{gb08a}.}
\end{minipage} 
\end{figure}
%

\subsection{Gravity torque maps: a new tool to quantify AGN fueling}

We have made a detailed case-by-case analysis of the distribution and kinematics of molecular gas in the galaxies of the core sample of NUGA, and interpreted these in terms of evidence of ongoing feeding. 
Our case-by-case in-depth study is the approach adopted to take advantage of the high quality of the NUGA maps\footnote{A statistical analysis of the whole $\sim$25 galaxy sample is currently underway}.
The gas response to the stellar potential is characterized with the help of high-resolution optical and NIR images of the galaxies. More precisely, we quantify the efficiency of the stellar potential to drain the gas angular momentum in a particular galaxy by deriving the gravity torques exerted by the potential on the gas \cite{gb05}. 
With this method we can look for the feeding agent in the stellar potential.

The results obtained from the analysis of stellar torques have revealed a puzzling feeding budget in the circumnuclear disks of the first NUGA targets studied by \cite{gb05}. Paradoxically, feeding due to the stellar potential seems to be presently inhibited close to the four AGNs analyzed, as torques are seen to be positive at 
$r<200$~pc. To solve the puzzle, \cite{gb05} suggest that gravity torques could be assisted by other mechanisms that become competitive in due time with non-axisymmetric perturbations. The authors estimate on a case-by-case basis that the gravity torque barrier associated with the ILRs of the galaxies analyzed in their paper could be overcome by viscosity (a source of negative torques), provided that the bar strength is moderate-to-low. Gravity torques and viscosity could thus combine their efforts to produce recurrent episodes of activity during the typical lifetime of any galaxy, a situation reminiscent of a self-regulation scenario, commonly invoked to understand the laws governing star formation and gravitational instabilities in galaxy disks. The self-destruction of bars, triggered by the radial re-distribution of gas and the implied angular momentum exchanges, illustrated through numerical simulations \cite{bou02, bou05}, may have important side-effects for AGN fueling: several AGN feeding episodes may occur during a single bar cycle \cite{gb05}.

The existence of short-lived (lifetime~$\leq$~10$^6$--10$^7$yr) feeding agents in the stellar potential (such as nuclear bars, oval perturbations or $m=1$ instabilities), may also explain the above mentioned gravity torque puzzle. In particular, this solution would account for the low chances of finding a smoking gun evidence of fueling, especially if the number of targets scrutinized is small \cite{gb05}.

As recently shown \cite{hun08, gb08a}, by analyzing the cases of the NUGA galaxies NGC~2782 and NGC~4579, to find a galaxy {\it caught in the act} of fueling the AGN is unlikely but not impossible, however. The disk of NGC~2782, analyzed by \cite{hun08}, has two embedded bar-like perturbations: a primary large-scale oval (of 5~kpc diameter) and a prominent nuclear bar (of 1.5~kpc diameter).  The decoupling of the nuclear bar has made it possible to transport molecular gas inside the ILR of the primary bar and then fuel the nuclear starburst and possibly the AGN. N-body simulations predict the decoupling of the nuclear bar and the ensuing gas inflow down to $r\sim$200~pc (Fig.~\ref{n2782}). More compelling, in terms of AGN feeding evidence, is the case of NGC~4579, recently revisited in \cite{gb08a}. NGC~4579 has a prominent bar+spiral structure in the outer disk, $r\geq$2~kpc, as well as an oval distortion in the inner $r\sim$200~pc. The gravity torque budget in NGC~4579, derived by \cite{gb08a}, shows that the fueling process is at work on different spatial scales (Figs.~\ref{n4579-1} and \ref{n4579-2}). In the outer disk, the decoupling of the spiral structure produces net gas inflow on intermediate scales. Most remarkably, the co-rotation barrier seems to be overcome due to secular evolution processes, a result found in other NUGA galaxies \cite{haa08b}. The gas in the inner disk of NGC~4579 is being funneled down to $r\sim$300~pc.  Closer to the AGN ($r<$200~pc), gas feels negative torques due to the combined action of the large-scale bar and the inner oval. The two $m=2$ modes act in concert to produce net gas inflow down to $r\sim$50~pc, providing a clear {\it smoking gun} evidence of fueling with associated short dynamical time-scales ($\sim$1--3 rotation periods).

The future instrumentation, of which the ALMA interferometer is a good example, will allow us to improve the statistics of these type of AGN surveys, by increasing the sizes of galaxy samples by orders of magnitude. In addition, the spatial resolutions typically reached by ALMA will provide a sharp view of the distribution and kinematics of molecular gas in the central pc regions of many nearby AGNs. Finally, we will also be able to analyze in detail the bar/AGN feeding cycles of distant galaxies and therefore compare fueling mechanisms as a function of redshift.

\section{The feedback of activity in galaxies}

The use of tracers specific to the dense molecular gas phase is a key to probe the feedback of activity on the interstellar medium of galaxies in the local universe. With typical densities $n(H_2)\geq$10$^{4-5}$cm$^{-3}$), the dense molecular gas phase is the component directly involved in the fueling of star formation and AGN episodes. The onset of activity in its different varieties (starburst or AGN) can cause the injection of vast amounts of energy into the molecular gas reservoirs of galaxies through strong radiation fields and massive gas flows. The availability of high-spatial resolution observations of molecular species other than CO makes it possible to use the study of molecular gas chemistry as a tool to track down galaxy evolution. In this context, it is worth noting that extragalactic chemistry can put major constraints on the predictions of current chemical models of molecular clouds. For historical reasons, these models have been confronted thus far to observations of molecular clouds in our Galaxy. Active galaxies can drive chemical complexity to a higher degree and also on larger scales compared our Galaxy, however. More important, extragalactic chemistry studies are a key to constrain conversion factors, which are required to derive gas masses from line luminosities. This is mostly relevant considering that the different phenomena associated to activity are suspected to heavily influence the excitation and/or the chemistry of many of the tracers of the dense molecular gas phase which are routinely used in extragalactic research. This information is crucial for any quantitative assessment of the efficiency of the star formation process in galaxies.


\subsection{Molecular gas chemistry in starbursts}

 Star formation can drive a large variety of chemical environments. Different processes can shape the evolution of molecular gas along the evolutionary track of a starburst: large-scale shocks, strong UV-fields, cosmic-rays, and X-rays. These processes are expected to be at work at different locations in the disk and/or the disk-halo interface of a starburst. During the fueling stage, gravity torques produced by density waves (bars and spirals) can funnel the gas inwards into the circumnuclear regions of galaxies. During this process, large-scale shocks can be generated in molecular gas. Furthermore, strong UV-fields are associated with the numerous HII regions of a typical starburst. The influence of UV fields in the surrounding medium can produce the propagation of Photon Dominated Region (PDR) chemistry in molecular gas disks. In a more advanced stage of a typical starburst episode, different types of chemistry can propagate into the halo of galaxies, entrained by massive galactic outflows.

Different strategies can be adopted to tackle extragalactic chemistry. Single-dish line surveys allow the frequency coverage in the search of complex molecules to be expanded by significant factors and can thus provide us with a global (low-spatial resolution) view of the molecular gas inventory of a starburst. The single-dish line survey of the prototypical starburst galaxy NGC~253 is a good illustration of this technique \cite{mar03,mar05,mar06}.

The use of interferometers, with their high spatial resolution capabilities, makes possible to make a giant step forward in the field. The examples of M~82 and IC~342, two case study starbursts extensively observed with the PdBI and the Owens Valley Radio Observatory (OVRO) interferometers, are paradigmatic in this respect. Virtually all of the large-scale SiO emission detected in the PdBI map of M~82 \cite{gb01} traces the disk-halo interface of the galaxy where episodes of mass injection are building up the gaseous halo. In contrast, widespread emission of the formyl radical, HCO, mapped in M~82 with the PdBI, reveals the propagation of PDR chemistry inside the disk of this starburst \cite{gb02}. As such, the markedly different distributions of SiO and HCO in M~82 revealed by the PdBI, illustrates the power of mm-interferometry to unveil the strong chemical differentiation present in the molecular gas disks of starbursts. The scenario of a giant PDR in the disk of M~82 has received further observational support from the high abundances derived for molecular tracers which are specific to PDR environments such as CN, HOC$^+$, CO$^+$ \cite{fue05,fue06}.
Maps recently obtained in CN and HOC$^+$ lines with the PdBI are being used to complete this modelling effort \cite{fue08,gb08b}. In the case of IC~342, marked differences in morphology are also revealed among the different molecular maps obtained \cite{mei05, use06}, using the OVRO and PdBI interferometers. Whereas the emission of tracers of PDR chemistry, like C$_2$H and C$^{34}$S, originates exclusively from the inner 50-100~pc circumnuclear region, where UV fields are high, a different group of molecules, including SiO, CH$_3$OH, and HNCO, is correlated with the expected location of large-scale shocks induced by the bar potential in the outer disk.


\begin{figure}[tb!]
\begin{center}
\includegraphics[angle=-90,width=35pc]{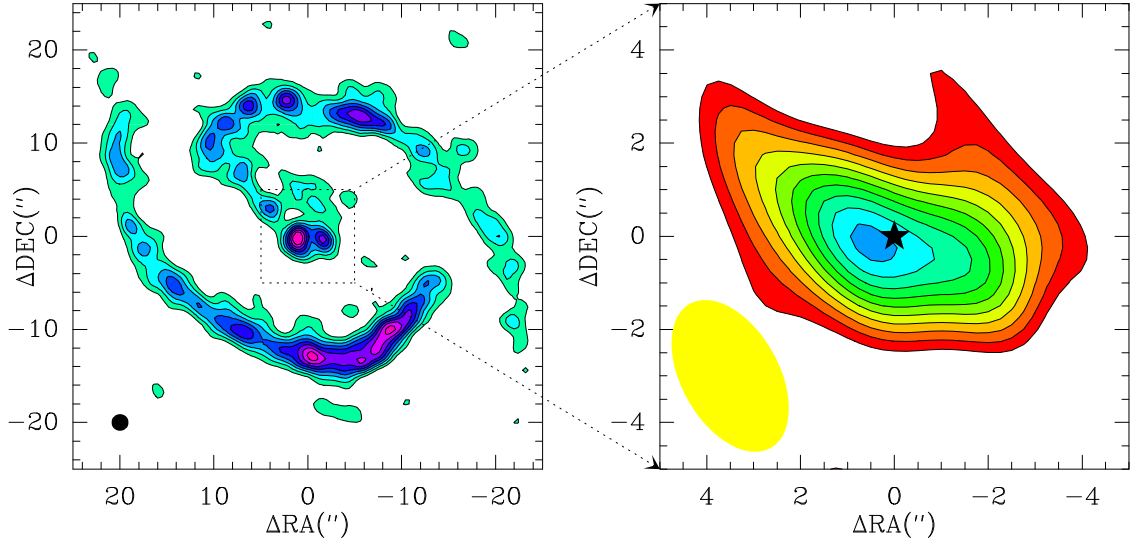}
\caption{\label{n1068-1} {\it(Left panel)} The CO(1--0) integrated intensity map of the Seyfert~2 galaxy NGC~1068 \cite{sch00} using the PdBI. {\it(Right panel)} Same for the SiO(2--1) line, but here showing the inner $r$\,$\sim$300~pc region of the circumnuclear molecular disk (CND) of the galaxy mapped with the PdBI. The filled ellipse represents the beam size. Figure adapted from \cite{gb08b}.}
\end{center}
\end{figure}



\begin{figure}[tb!]
\begin{center}
\includegraphics[angle=-90,width=\textwidth]{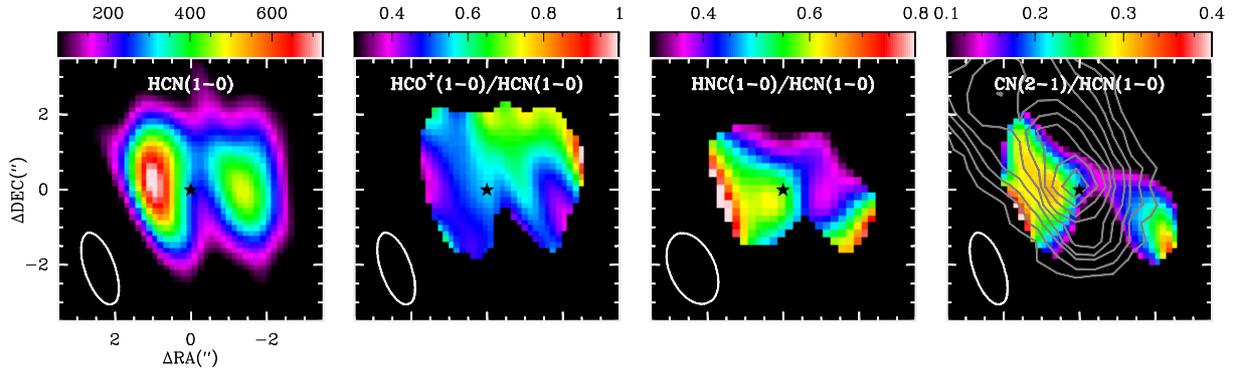}
\caption{\label{n1068-2} {\it(Left panel)} The HCN(1--0) line intensity map obtained with the PdBI towards the CND of NGC~1068. {\it (Middle)} and {\it (right panels)} The HCO$^+$/HCN, HNC/HCN and CN/HCN line ratio maps of the CND of NGC~1068 derived from PdBI data. Contours in the right panel stand for X-ray emission mapped by Chandra from 0.25 to 7.50~keV \cite{you01}. Figure adapted from \cite{use08}} 
\end{center}
\end{figure}


\subsection{Molecular gas chemistry in AGNs}

Based on the observation of local template galaxies, it has been suggested that the excitation and/or the chemistry of molecular gas can show significant differences between starbursts and AGNs \cite{koh01, kri08}. A recent study \cite{kri08} has shown that starbursts and AGNs populate different regions in a set of diagnostic diagrams that make use of several HCN and HCO$^+$ line ratios. In particular, Large Velocity Gradient (LVG) fits to the observed ratios indicate lower H$_2$ densities, and larger HCN abundances in AGNs compared to starburst galaxies. 
Relative to other tracers of the dense molecular gas, overluminous HCN lines seem to be common in the circumnuclear disks (CND) of many Seyfert galaxies \cite{tac94, koh01, use04, kri08}. It is tempting to link the existence of  overluminous HCN lines with the influence on molecular gas of some of the distinctive ingredients of AGN activity. 
Among these, X-rays are suspected to be able to process large column densities of molecular gas around AGNs, producing X-ray dominated regions (XDR). The abundances of certain ions, radicals and molecular species can be enhanced \cite{lep96, mal96, meij05, meij07}. In particular, and while still a highly debated issue, it has been discussed that the abundance of HCN can be boosted in XDRs, partly accounting for the reported observational dichotomy between starbursts and AGNs. Alternatively, the excitation of HCN lines in AGNs, some of them being sources with a high a infrared luminosity, might be affected by IR pumping through a 14$\mu$m vibrational transition of the molecule \cite{aal95,gb06}. 

The large HCN abundances measured in the nucleus of the Seyfert 2 galaxy NGC~1068 were the first observational evidence that molecular gas chemistry can be influenced by AGN activity \cite{tac94, ste94}. The IRAM 30m telescope was used \cite{use04} to make complementary observations of the CND of NGC\,1068 in eight molecular species. The global analysis of the survey, which includes several lines of SiO, CN, HCO, H$^{13}$CO$^{+}$, H$^{12}$CO$^{+}$, HOC$^{+}$, HCN, CS, and CO, suggests that the bulk of the molecular gas emission in the CND of NGC\,1068 can be interpreted as coming from a giant XDR. Recently, a high-resolution PdBI map has provided a sharp view of the distribution of the SiO emission coming from the CND of NGC~1068 \cite{gb08c} (Fig.~\ref{n1068-1}). The enhancement of SiO in the CND of NGC~1068 cannot be attributed to the action of ongoing star formation as there is negative evidence of a recent starburst in the circumnuclear region of this Seyfert \cite{dav07}. Alternatively, the processing of 10~\AA~dust grains by X-rays, as a mechanism to enhance silicon chemistry in gas phase, might explain the large SiO abundances of the CND. The inclusion of dust grain chemistry could solve the controversy between different gas-phase XDR schemes regarding the abundance of some molecular tracers like HCN in X-ray irradiated environments. New maps of NGC~1068 obtained with the PdBI in molecular species like HCN, HCO$^+$, HNC and CN are being used to study the chemical differentiation inside the CND of the galaxy, and explore its dependence with the illumination of molecular gas by the strong X-ray source of the nucleus \cite{use08} (Fig.~\ref{n1068-2}).

\subsection{Molecular gas chemistry in IR luminous galaxies}

The use of HCN lines as a quantitative tracer of the dense molecular gas in galaxies can be questioned if the excitation conditions and/or the chemical environment of molecular gas depart from normality. This can alter the conversion factor between the luminosity of HCN lines and the mass of dense molecular gas. The doubts cast on HCN call for the observation of alternative tracers of dense gas in luminous and ultraluminous infrared (IR) galaxies (LIRGs and ULIRGs). This question is crucial to disentangle the power sources of the infrared luminosities of these galaxies, where both star formation and AGN activity are embedded.

A recent paper \cite{gc08} has presented evidence that the $L_{\rm FIR}/L_{\rm HCN(1-0)}$ ratio, taken as a proxy for the star formation efficiency of the dense gas (SFE$_{\rm dense}$), is a factor $\sim$2--3 higher in IR luminous galaxies ($L_{\rm FIR} > 10^{11}\,L_{\odot}$) compared to normal galaxies. Local universe LIRGs and ULIRGs populate a region in the SFE$_{\rm dense}$ diagram that lies between those occupied by normal and high-$z$ IR luminous galaxies. The reported trend in the SFE$_{\rm dense}$ derived from HCN data implies that there is a statistically significant turn upward in the Kennicutt-Schmidt law, $\Sigma_{\rm SFR} \propto \Sigma_{\rm dense}^{N}$, at high $L_{\rm FIR}$: $N$ changes from $\sim$0.80--0.95 (for $L_{\rm FIR} < 10^{11}\,L_{\odot}$) to $\sim$1.1--1.2 (for $L_{\rm FIR} > 10^{11}\,L_{\odot}$). These results confirm the predictions of star formation models \cite{kru07}. In addition, and based on a multi-line LVG analysis of HCN and HCO$^{+}$ data that follows the survey published in \cite{gc06}, \cite{gc08} find that the the conversion factor between $L_{\rm HCN(1-0)}$ and the mass of dense molecular gas is $\sim$3 times lower at high $L_{\rm FIR}$. Both factors taken into account, this result reinforces a scenario where the SFE$_{\rm dense}$ could well be up to an order of magnitude higher in extreme LIRGs/ULIRGs compared to normal galaxies. Of particular note, a significant overabundance of HCN has also been reported in the IR luminous z$\sim$4 quasar APM~08279 \cite{wag05, gb06}, an example of extreme chemistry and excitation conditions of molecular gas at high redshift.




\ack
S. Garc{\'{\i}}a-Burillo (SGB) would like to thank the organisers of AHAR-2008 for inviting him to this fruitful conference. SGB would like to thank all the members of the NUGA team, as well as all his collaborators on the study
of the chemistry of molecular gas in galaxies, presented in this paper. We all acknowledge the IRAM staff from the Plateau de Bure, from Grenoble and from Pico Veleta for carrying out the observations and help provided during the data reduction.


\section*{References}
\begin{thebibliography}{60}


\bibitem{kna00} Knapen J~H, Shlosman I and Peletier R~F\ 2000 {\it Astrophys. J.} {\bf 529} 93 

\bibitem{com03} Combes F\ 2003 {\it Active Galactic Nuclei: From Central Engine to Host Galaxy} vol 290 p 411 

\bibitem{mart03} Martini P, Regan M~W, Mulchaey J~S and Pogge R~W\ 2003 {\it Astrophys. J.} {\bf 589} 774 

\bibitem{mart04} Martini P\ 2004 {\it The Interplay Among Black Holes, Stars and ISM in Galactic Nuclei. IAU-222} vol 222, p 235 

\bibitem{wad04} Wada K\ 2004 {\it Coevolution of Black Holes and Galaxies} p 186

\bibitem{kin07} King A~R and Pringle J~E\ 2007 {\it Mon. Not. Roy. Astron. Soc.} {\bf 377} L25 

\bibitem{sel04} Sellwood J~A and Shen J\ 2004 {\it Coevolution of Black Holes and Galaxies} p 203 


\bibitem{gb03a} Garc{\'{\i}}a-Burillo S et al.\ 2003a {\it Active Galactic Nuclei: From 
Central Engine to Host Galaxy} vol 290 p 423 

\bibitem{gb03b} Garc{\'{\i}}a-Burillo S et al.\ 2003b {\it Astron. Astrophys.} {\bf 407} 485 

\bibitem{haa08a} Haan S, Schinnerer E, Mundell C~G, Garc{\'{\i}}a-Burillo S and Combes F\ 2008a  {\it Astron. J.} {\bf 135} 232

\bibitem{haa08b} Haan S, Schinnerer E, Emsellem E, Garc{\'{\i}}a-Burillo S, Combes F, Mundell C~G and Rix H-W \ 2008b {\it Astrophys. J.} submitted

\bibitem{gb08a} Garc{\'{\i}}a-Burillo S et al.\ 2008a {\it Astron. Astrophys.} submitted 

\bibitem{kri05} Krips M et al.\ 2005 {\it Astron. Astrophys.} {\bf 442} 479 

\bibitem{com04} Combes F et al.\ 2004 {\it Astron. Astrophys.} {\bf 414} 857 

\bibitem{cas08a} Casasola V, Combes F, Garcia-Burillo S, Hunt L~K, Leon S and Baker A~J\ 2008a {\it Astron. Astrophys.}, ArXiv e-prints, 808, arXiv:0808.1186 

\bibitem{cas08b} Casasola V et al \ 2008b {\it Astron. Astrophys.} in preparation

\bibitem{boo07} Boone F et al.\ 2007 {\it Astron. Astrophys.} {\bf 471} 113 

\bibitem{hun08} Hunt L~K et al.\ 2008 {\it Astron. Astrophys.} {\bf 482} 133
 
\bibitem{lin08} Lindt-Krieg E, Eckart A, Neri R, Krips M, Pott J-U, Garc{\'{\i}}a-Burillo S and Combes F\ 2008 {\it Astron. Astrophys.} {\bf 479} 377 


\bibitem{gb05} Garc{\'{\i}}a-Burillo S, Combes F, Schinnerer E, Boone F and Hunt L~K\ 2005 {\it Astron. Astrophys.} {\bf 441}, 1011 

\bibitem{kri07} Krips M et al.\ 2007 {\it Astron. Astrophys.} {\bf 468} L63 

\bibitem{bou02} Bournaud F and Combes F\ 2002 {\it Astron. Astrophys.} {\bf 392} 83 

\bibitem{bou05} Bournaud F, Combes F and Semelin B\ 2005 {\it Mon. Not. Roy. Astron. Soc.} {\bf 364} L18 

 
\bibitem{mar03} Mart{\'{\i}}n S, Mauersberger R, Mart{\'{\i}}n-Pintado J, Garc{\'{\i}}a-Burillo S and Henkel C\ 2003 {\it Astron. Astrophys.} {\bf 411} L465 

\bibitem{mar05} Mart{\'{\i}}n S, Mart{\'{\i}}n-Pintado J, Mauersberger R, Henkel C 
and Garc{\'{\i}}a-Burillo S\ 2005 {\it Astrophys. J.} {\bf 620} 210 

\bibitem{mar06} Mart{\'{\i}}n S, Mauersberger R, Mart{\'{\i}}n-Pintado J, Henkel C and Garc{\'{\i}}a-Burillo S\ 2006 {\it Astrophys. J. Suppl.} {\bf 164} 450 

\bibitem{gb01} Garc{\'{\i}}a-Burillo S, Mart{\'{\i}}n-Pintado J, Fuente A and Neri R\ 2001 {\it Astrophys. J. Lett.} {\bf 563} L27 

\bibitem{gb02} Garc{\'{\i}}a-Burillo S, Mart{\'{\i}}n-Pintado J, Fuente A, Usero A and Neri R\ 2002 {\it Astrophys. J. Lett.} {\bf 575} L55 

\bibitem{fue05} Fuente A, Garc{\'{\i}}a-Burillo S, Gerin M, Teyssier D, Usero A, Rizzo J~R and de Vicente P\ 2005 {\it Astrophys. J. Lett.} {\bf 619} L155 

\bibitem{fue06} Fuente A, Garc{\'{\i}}a-Burillo S, Gerin M, Rizzo J~R, Usero A, Teyssier D, Roueff E and Le Bourlot J\ 2006 {\it Astrophys. J. Lett.} {\bf 641} L105 

\bibitem{fue08} Fuente A et al.\ 2008 {\it Astron. Astrophys.} submitted  

\bibitem{gb08b} Garc{\'{\i}}a-Burillo S\ 2008b {\it Astrophys. J.} in preparation

\bibitem{mei05} Meier D~S and Turner J~L\ 2005 {\it Astrophys. J.} {\bf 618} 259 

\bibitem{use06} Usero A, Garc{\'{\i}}a-Burillo S, Mart{\'{\i}}n-Pintado J, Fuente A and Neri R\ 2006 {\it Astron. Astrophys.} {\bf 448} 457 

\bibitem{koh01} Kohno K, Matsushita S, Vila-Vilar{\'o} B, Okumura S~K, Shibatsuka T, Okiura M, Ishizuki S and Kawabe R\ 2001 {\it The Central Kiloparsec of Starbursts and AGN: The La Palma Connection} vol 249 p 672 

\bibitem{kri08} Krips M, Neri R, Garc{\'{\i}}a-Burillo S, Mart{\'{\i}}n S, Combes F, Graci{\'a}-Carpio J. and Eckart A\ 2008  {\it Astrophys. J.} {\bf 677} 262 

\bibitem{tac94} Tacconi L~J, Genzel R, Blietz M, Cameron M, Harris A~I and Madden S\ 1994 {\it Astrophys. J. Lett.} {\bf 426} L77

\bibitem{use04} Usero A, Garc{\'{\i}}a-Burillo S, Fuente A, Mart{\'{\i}}n-Pintado J and Rodr{\'{\i}}guez-Fern{\'a}ndez N~J\ 2004 {\it Astron. Astrophys.}  {\bf 419} 897 

\bibitem{lep96} Lepp S and Dalgarno A\ 1996 {\it Astron. Astrophys. Lett.} {\bf 306} L21 

\bibitem{mal96} Maloney P~R, Hollenbach D~J and Tielens A~G~G~M\ 1996 {\it Astrophys. J.} {\bf 466} 561 

\bibitem{meij05} Meijerink R and Spaans M\ 2005 {\it Astron. Astrophys.} {\bf 436} 397 

\bibitem{meij07} Meijerink R, Spaans M and Israel F~P\ 2007 {\it Astron. Astrophys.} {\bf 461} 793 

\bibitem{aal95} Aalto S, Booth R~S, Black J~H and Johansson L~E~B\ 1995 {\it Astron. Astrophys.} {\bf 300} 369 

\bibitem{gb06} Garc{\'{\i}}a-Burillo S et al.\ 2006  {\it Astrophys. J. Lett.} {\bf 645} L17 

\bibitem{ste94} Sternberg A, Genzel R and Tacconi L~J\ 1994  {\it Astrophys. J. Lett.} {\bf 436} L131 

\bibitem{gb08c} Garc{\'{\i}}a-Burillo S\ 2008c {\it Astrophys. J.} in preparation

\bibitem{sch00} Schinnerer E, Eckart A, Tacconi L~J, Genzel R and Downes D\ 2000 {\it Astrophys. J.} {\bf 533} 850 

\bibitem{dav07} Davies R~I, Mueller-S{\'a}nchez F, Genzel R, Tacconi L~J, Hicks E~K~S, Friedrich S and Sternberg A\ 2007 {\it Astrophys. J.} {\bf 671} 1388 

\bibitem{use08} Usero A et al. 2008 {\it Astron. Astrophys.} in preparation 

\bibitem{you01} Young A~J, Wilson A~S and Shopbell P~L\ 2001 {\it Astrophys. J.} {\bf 556} 6 

\bibitem{gc08} Graci{\'a}-Carpio J, Garc{\'{\i}}a-Burillo S, Planesas P, Fuente A and Usero A\ 2008 {\it Astron. Astrophys.} {\bf 479} 703 

\bibitem{kru07} Krumholz M~R and Thompson T~A\ 2007 {\it Astrophys. J.} {\bf 669} 289 

\bibitem{gc06} Graci{\'a}-Carpio J, Garc{\'{\i}}a-Burillo S, Planesas P and Colina L\ 2006 {\it Astrophys. J. Lett.} {\bf 640} L135 

\bibitem{wag05} Wagg J, Wilner D~J, Neri R, Downes D and Wiklind T\ 2005 {\it Astrophys. J. Lett.} {\bf 634} L13 

\end{thebibliography}
\end{document}